\documentclass[pra,twocolumn,letterpaper,noeprint,longbibliography,nodoi,footinbib]{revtex4-1} 
\usepackage[colorlinks, allcolors=blue]{hyperref}

\usepackage{amsmath}  
\usepackage{amsfonts} 
\usepackage{graphicx} 
\usepackage{xcolor}
\usepackage{tikz}

\newcommand{\be}{\begin{equation}}
\newcommand{\ee}{\end{equation}}
\newcommand{\ba}{\begin{align}}
\newcommand{\ea}{\end{align}}

\newif\ifredlined

\redlinedfalse

\ifredlined
\newcommand\circled[1]{\marginpar{\revcolor{red}\tikz[baseline=(char.base)]{\node[shape=circle,draw,inner sep=2pt] (char) {#1};}}}
\newcommand\revcolor[1]{\color{#1}}
\else
\newcommand\circled[1]{}
\newcommand\revcolor[1]{}
\fi

\begin{document}

\title{Exact energy-eigenstates of the Coulomb-Stark Hamiltonian}


\author{S. Yusofsani}
\author{M. Kolesik}
\affiliation{College of Optical Sciences, University of Arizona, Tucson, AZ 85721, USA}

\date{\today}

\begin{abstract}
  An approximation-free, numerically efficient algorithm is presented for
  the Hamiltonian eigenstates of the Stark-Hydrogen problem describing a
  quantum particle exposed to the central Coulomb force and a homogeneous external field.
  As an example of application in a state-expansion with continuous energy,
  we calculate the time-dependent wave function of an electron tunneling from a
  hydrogen atom suddenly exposed to an external electric field.
\end{abstract}

\maketitle
\section{Introduction}
In calculating the time evolution of a pure quantum state, the most fundamental approach is to expand the
wave function in terms of the eigenstates of the Hamiltonian. While this is relatively straightforward 
in finite-dimensional Hilbert spaces, infinite-dimensional cases are more difficult. 
Calculating the eigenstates of the system and {\em using them as a basis}
for the space of quantum states
is particularly challenging when
the Hamiltonian has a continuous spectrum. In such a situation, at least some of the eigenstates are not
normalizable in the usual sense, and normalization to the Dirac-delta function in energy is rather subtle
from the numerical standpoint. For this reason, in most practical situations one resorts to approximations,
for example by expanding solutions into approximate, ``discretized-continuum'' basis.

An important example of a system with a continuous spectrum is the Stark-Coulomb Hamiltonian~\cite{bookCycon}, i.e. the problem of a particle
subject to a constant external electric field and a central Coulomb force. Even an arbitrarily weak external field causes a
qualitative transformation of the original hydrogen-atom spectrum into infinitely degenerate continuum encompassing the
whole real axis. The zero-field discrete-energy states mutate into Stark resonances~\cite{emmanouilidou}  which further complicate
exact treatments. Perhaps not surprisingly, numerically exact calculation of the energy-eigenstates for this system
has not yet been reported. In this paper we present a robust, approximation-free method to calculate all the eigenstates
in a way that allows practical applications, including expansion of time-dependent wave functions.

Obviously, the Stark-Coulomb Hamiltonian is important for a number of applications. Recently, experiments 
on atomic hydrogen~\cite{SainadhTunnelTime} exposed to strong electric fields provide a vital testing ground for theory and
simulations~\cite{Kellerstreak,Keller13,Landsman14,ZimmermanTunnelTime,SainadhTunnelTime,SainadhTunnelTimeReveiw}. Moreover, in ionization of
atoms and molecules~\cite{StarkDFT}, electrons liberated from the neutral systems
experience the Stark-Coulomb potential at larger distances from the nucleus ~\cite{Tong,CiappinaCoulombStark,ZHANGBin}.
This composite potential affects
the properties of the  electronic wave functions contributing to strong-field ionization~\cite{Kolesik:19,Smirnova15,backpropzero,Ni18}, and high-harmonic generation~\cite{CiappinaCoulombStark}.
Thus, the numerically exact wave functions of such a problem can be of great practical use.

The rest of this paper is organized as follows: We start by laying out the theoretical foundations, by reviewing the
Stark-Coulomb Hamiltonian in parabolic coordinate system. We proceed by calculating the exact eigenstates
including their proper normalization to the Dirac-delta function in energy.
We then demonstrate the power of this method by using it to describe the time evolution of an electron
tunneling from the ground state of hydrogen after a sudden exposure to an external electric field.

\section{The Coulomb-Stark Problem}
We start with the formulation of the problem of a quantum particle (electron) in a Coulomb potential and subject to an
external electric field. It has been recognized that for the calculation of Stark resonances the 
parabolic coordinate system is the most suitable~\cite{Damburg_1978,Jannussis,HighlyA,BadDirect,Comment}.
Naturally, this is also the case for the continuum-energy eigenstates
we are interested in, and we shall use this frame of reference as well. In this section we recall the
equations which serve as our point of departure.

The coordinate relations between the parabolic coordinates $(u,v,\phi)$ and Cartesian coordinates $(x,y,z)$ are:
\begin{align}
z = (u-v)/2, \ \ \
x = (uv)^{1/2} \cos \phi, \ \ \  \\
\nonumber
y = (uv)^{1/2} \sin \phi,\ \ \ 
\phi = \tan^{-1} \frac{y}{x} \ \ \ \\ \nonumber \ \ \
u,v\in [0,\infty), \ \ \ \phi\in [0,2\pi) \ ,
\end{align}
and the volume element reads:
\be
dV = \frac{u+v}{4} du dv d\phi \ .
\ee
The main advantage of this coordinate system is that 
the Schr\" odinger equation of an electron in a Coulomb potential
of a singly-charged nucleus (hydrogen) remains separable even in the presence
of an external electric field.

The time independent Schr\"odinger equation (TISE) becomes:
\begin{align}
&H\Psi(u,v,\phi) = W \Psi(u,v,\phi) = \\
&-\!\frac{2}{u\!+\!v}\!\!
\left[
\frac{\partial}{\partial u}u\frac{\partial \Psi}{\partial u}\!+\!
\frac{\partial}{\partial v}v\frac{\partial \Psi}{\partial v}
\right]\!\!
-\!\frac{1}{2 u v} \frac{\partial^2\Psi}{\partial\phi^2}\!
-\!\frac{2\Psi}{u\!+\!v}\!-\!F\frac{u\!-\!v}{2}\Psi
\nonumber
\end{align}
where $F$ is the strength of the homogeneous external field.
Without any loss of generality we will assume that this quantity is positive.

As we will see in a moment, the energy eigenstates $\Psi_{qW}$ for this Hamiltonian can be labeled by two
kinds of quantum numbers; The first labels a continuum of energies, $W\in\mathbb{R}$,
encompassing the whole real axis, and the second is a composite discrete pair $q=\{n_v,m\}$
where $m$ is an integer standing for the usual ``magnetic'' quantum number, and $n_v$ is a whole
number counting the  zero-crossing along the parabolic axis $v$.

An arbitrary  time-dependent wave function can be expanded in this eigenenergy basis as:
\be
\Psi(t) = 
\sum_q \int dW A_q(W) e^{-iWt} \Psi_{qW}  \ ,
\ee
where $A_q(W)$ is the wave function in the energy representation, and it can be understood as the overlap
integral between the given quantum state (at time $t=0$) and the corresponding eigenstate.

Note that $\Psi_{qW}$ depend parameterically on $F$, so there is a continuum of different bases,
distinguished by the value of the external field. For each fixed $F$, we  have an infinitely degenerate
continuum of energies. Our task is to design an accurate and efficient numerical algorithm to
evaluate these states without any approximations.

Because of the symmetry, the sought after eigenfunctions depend on the azimuthal quantum number
as $\Psi\sim e^{i m \phi}$. The methods required for different $m$ are completely analogous,
so for the sake of simplicity we will restrict ourselves to the case $m=0$.
In order to further simplify our notation, we will use a shorthand, $n=\{n,m=0\}$,
for the discrete part of the eigenstate label.

Using the method of separation of variables we can turn the partial differential equation into
a pair of ordinary differential equations. With this in mind, 
we  rewrite the TISE as
\be
-\frac{\partial}{\partial u}u\frac{\partial \Psi}{\partial u}
-\frac{\partial}{\partial v}v\frac{\partial \Psi}{\partial v}
 -\! F\frac{u^2\!-\!v^2}{4}\Psi-W\frac{u\!+\!v}{2}
=\Psi \ ,
\ee
and note that it has the form
\be
\hat{h}_u\Psi +\hat{h}_v\Psi = \Psi
\ee
with
\begin{align}
\hat{h}_u
&=
-\frac{\partial}{\partial u}u\frac{\partial}{\partial u}
-\frac{Fu^2}{4}
-\frac{Wu}{2}
\\ 
\hat{h}_v
&=
-\frac{\partial}{\partial v}v\frac{\partial}{\partial v}
+\frac{Fv^2}{4}
-\frac{Wv}{2} \ .
\end{align}
We use the following ansatz
\be
\Psi_{nW} = V_{nW}(v) U_{nW}(u)
\label{eq:PsiUV}
\ee
leading to these separated equations for $U_{nW}(v)$
\be
\hat{h}_u
U_{nW}(u) = z_u(n,W,F)U_{nW}(u) \ ,
\label{eq:U}
\ee
 and $ V_{nW}(u)$:
\be
\hat{h}_v
V_{nW}(v) = z_v(n,W,F)V_{nW}(v)
\label{eq:UVeqs}
\ee
with the two separation constants tied by the constraint
\be
z_u(n,W,F)+z_v(n,W,F)=1 \ .
\label{eq:separation}
\ee
Note that the above equations are  well-known from the Stark-resonance problem
in atomic hydrogen~\cite{Damburg_1978}. Unlike in the discrete non-Hermitian resonance calculation, we look for a
continuum of standard, i.e. Hermitian, real-valued energy eigenstates. The distinction between these
two kinds of eigen-value problems boils down to the boundary conditions imposed on the
eigenfunction at infinity. While the frequently-studied Stark resonances with complex-valued energies must
behave as outgoing waves, the Hermitian eigenfunctions possessing real-valued energies behave as
standing waves at infinity. In other words, the main challenge here is not in finding the eigenvalues,
but in designing an approach to obtain the properly normalized wave function, and it is the continuum nature
of the spectrum that makes the problem difficult.

An important aspect of
this work is that our approximation-free numerical calculation is sufficiently efficient and accurate
at the same time, so that it makes it possible to use the resulting continuum-energy basis
for state expansion of arbitrary wave functions.

\section{Calculation of  eigenstates}

The most challenging step in the calculation of the wave functions corresponding to a
continuum of energies is to ensure the correct normalization for $U_{nW}$. 
For applications in which the set of eigenstates is used as a basis, it is necessary that the resolution
of unity in the energy space is achieved. This will require
construction of the inner and outer solutions which smoothly connect to each other,
each supplying a crucial piece of information.
We take the following steps to calculate $U_{nW}$ with the correct normalization factor.

\textbf{Inner solution via ``analytic continuation'':}
Both $V_{nW}(u)$ and  $U_{nW}(u)$ can be easily obtained in the form of a series  in the vicinity of
an arbitrary point, say $u=a$, provided that the function value and its derivative are known at this
point. Utilizing a large number of terms in such a series, such initial data can be obtained at
$u=a + \delta a$, and the series expansion can be re-done around this new ``center.'' This is akin
to the analytic continuation of holomorphic functions of a complex variable. The process can be
extended to an arbitrary distance from the origin, but it leaves the normalization
of  $U_{nW}(u)$ undetermined.

\textbf{Outer solution via carrier-envelope method:}
In order to fix the normalization, we examine the large-$u$ behavior of the differential equation
for $U$ and split the function into a ``carrier-wave'' and  its slowly changing ``envelope.''
Such a form allows us to ensure the correct normalization to a Dirac-delta function in energy.
However, it will not fix the relative phase-shift between the incoming and outgoing waves.

\textbf{Joining the inner and outer solutions:}
Because both representations solve the same differential equation,
the remaining degrees of freedom, namely the normalization of the inner part
and the phase shift of the outer part can be found by requiring that the two
agree at two arbitrarily chosen points.

Before we execute this plan, it should be useful to point out some
qualitative differences between  $V_{nW}(v)$ and  $U_{nW}(u)$, so we
start with their spectral properties next.

\subsection{Discrete and continuous spectra and the eigenstate normalization}

One can see that for large values of $u$ and $v$ the equations for $U$ and $V$ will turn into:
\begin{equation}
-\frac{1}{2}\frac{\partial^2}{\partial u^2}U
-\frac{F}{8} u U
\approx \frac{W}{4} U \phantom{\ .}
\label{eq:asymEqsU}
\end{equation}
and
\begin{equation}
-\frac{1}{2}\frac{\partial^2}{\partial v^2}V
+\frac{F}{8}v V
\approx \frac{W}{4} V \ .
\label{eq:asymEqsV}
\end{equation}
These equations  resemble the TISE of a particle living on a half-line with a potential that 
is pulling it from the origin (in the equation for $U$) and/or pushing it towards the origin (in the equation
for $V$, which is sometimes called the ``quantum bouncer''~\cite{qbouncer}).

Hence the spectrum of $U$ is continuous while the spectrum of $V$ is discrete in the following sense.
For any given real value of $W$ (with a fixed $F$) there is a discrete (infinite) set of eigenvalues
$z_v(n,F,W)$. In contrast, for a given $z_u = 1 - z_v$, the equation for $U$ has a delta-normalizable
solution for any $W\in\mathbb{R}$. In other words, the normalization for $V_{nW}$s is to ``Kronecker delta,"
while the normalization of $U_{nW}$s  is to ``Dirac delta." 

The nature of the spectrum is reflected in the following normalization relation
which the eigenstates must obey, 
\begin{align}
\langle\Psi_{nW}|\Psi_{mE}\rangle \nonumber
&=\\  \nonumber
2\pi\int \frac{u+v}{4} 
& V_{nW}(v) V_{mE}(v)
U_{nW}(u) U_{mE}(u)
dvdu\\ 
&=
\delta_{nm}\delta(W-E) \ .
\label{eq:complete}
\end{align}
This in turn implies the  orthogonality relations between the $U$ and $V$ functions:
\begin{align}
&\int 
V_{nW}(v) V_{mW}(v)
dv
=
\delta_{nm} \ ,
\label{eq:Vort}
\\
&\frac{\pi}{2}\int 
u \ U_{nW}(u)  U_{nE}(u)
du
=\delta(W-E) \ .
\label{eq:norms}
\\
\text{and} \  \ &\int
U_{nW}(u) U_{mE}(u)
du
=0 \ .
\end{align}
It is the second, Dirac-delta normalization condition that requires more attention
if one aims for a practical method to utilize 
these states in expansion of arbitrary wave functions.

  The third condition holds in the distribution sense and is actually independent
  of the normalization.
  While it may seem that the term proportional to $v$ in the normalization integral (\ref{eq:complete})
  was left out to obtain (\ref{eq:norms}), one must evaluate all wave function overlaps
  with the whole volume element proportional to $(u+v)/4$. The detailed derivation
  of the above normalization conditions is given in the Appendix.

  For readers who may prefer a more intuitive normalization argument, we note that these conditions
  may be obtained by following the recipe given in Refs.\cite{Landau,LucKoenig1}.
  This approach splits the wave function into the incoming and outgoing waves, and requires that
  the total probability outflow in the latter equals $1/2\pi$ for all continuum energies.
  It is straightforward to demonstrate that the above normalization satisfies this requirement.

\subsection{$V_{nW}$ eigenstates}

\begin{figure}[h]
  \includegraphics[clip,width=0.8\linewidth]{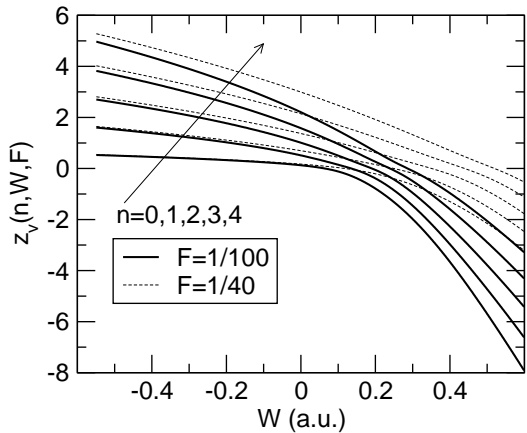}
\caption{Separation constant $z_v(n,W,F)$ as a function of energy  $W$
for $n=0,1,2,3,4$ (indicated by arrow) and two values of the external field.}
\label{fig:zv}
\end{figure}

While for $F>0$ closed-form expressions for the $V_{nW}$ solutions to~(\ref{eq:UVeqs}) are not known, they are normalizable
and can be easily calculated using the method of series expansion.
We will use the following representation
\be
V_{nW}(v) = v_{n0}(W,F)M(W,z_v,F|v) \ ,
\label{eq:VEq}
\ee
in which $M$ is  a series in $v$ normalized such that $M(W,z,F|v=0)=1$ and $v_{n0}$ represents
the first coefficient in the series expansion for a normalized eigenstate.
(note that
  argument $z$ of $M$ here and in what follows stands for the separation constant $z_v$ or $z_u$,
  not a coordinate).
The series expansion can be written as:
\be
M(W,z,F|v) = \sum_{k=0}^\infty c_k(W,z,F)v^k \ ,
\label{eq:MVEq}
\ee
and substituting this in the differential equation we obtain
\begin{align}
\nonumber
&c_0 = 1,\\ \nonumber
&c_1 = -z,\\ \nonumber
&c_2 = \frac{1}{8}(2z^2-W),\\
&c_k(W,z,F) = 
\frac{1}{k^2}\left(\frac{F}{4}c_{k-3}-\frac{W}{2}c_{k-2}-zc_{k-1}\right)\ .
\label{eq:VcoeffEq}
\end{align}
Using this recurrence relation, hundreds of terms can be calculated efficiently.
Note that this series expansion does not represent the solution to our problem until the separation constant $z=z_v$
is determined. For a general $z$, the above function diverges at infinity, while we seek a normalizable $V_{nW}(v)$.
We can numerically calculate $z_v$ by demanding that the series expansion approximation remains reasonably close to
zero at a sample $v_0$ far enough from the origin and adjusting $z_v$ gradually until the function tends to zero
for large arguments $v$. In this process, one obtains a function which can have none or several zeros. The number
$n$ of these nodes is the first part of the discrete label of the energy eigenstate.
At this point we have determined
\be
z_v = z_v(n,W,F) \ \ n=0,1,\ldots
\ee
for any fixed $W$ and positive $F$. The behavior of $z_v$ as a function of energy $W$ is
illustrated in Fig.~\ref{fig:zv}.

\smallskip
\noindent

\begin{figure}[h]
  \includegraphics[clip,width=0.8\linewidth]{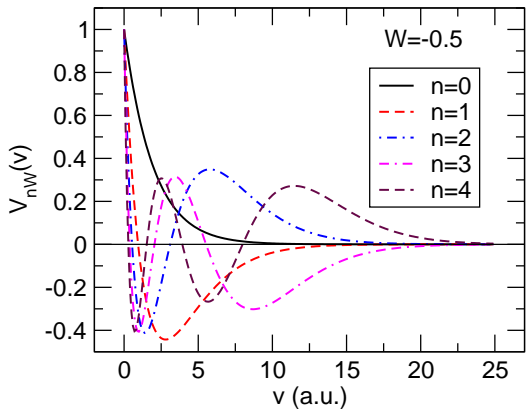}
\caption{Unnormalized function $V_{nW}(v)$ for several values of $n$
and $F=1/40$.}
\label{fig:Vns}
\end{figure}

Having fixed $z_v(n,W,F)$, formulas (\ref{eq:MVEq}) and (\ref{eq:VcoeffEq}) 
provide the shape of the function $V_{nW}(v)$. Figure~\ref{fig:Vns}
shows a few as an example.

\smallskip
\noindent

Finally, we determine the value $v_{n0}$ by imposing the normalization
condition~(\ref{eq:Vort}).
  The integrals required for this calculation can be obtained
  analytically in terms of the series-expansion coefficients, so $v_{n0}$
  can be highly accurate.

From the numerical standpoint, calculation of the $V$-part of the
energy eigenstate is straightforward, and various algorithms can be utilized
to find $z_v$ to a high accuracy. For the illustrations in this work,
we utilized between hundred and two hundred terms for the
series expansion around $v=0$. This is sufficient to reach
$v\approx 40 - 50$ with sufficient accuracy. Should $V_{nW}(v)$ be
needed for even larger $v$, the technique described next can also be used.

\subsection{$U_{nW}$ eigenstates}
The more challenging part of the problem is to calculate the $U$ eigenfunctions which are not normalizable
and have a continuous spectrum. After we determined the $z_v$ using the properties of the $V$-functions,
we can use the relation between $z_v$ and $z_u$ eigenvalues and calculate $z_u = 1-z_v$.

The behavior of $U$ at small $u$ can be calculated using the same method as the $V$ eigenfunctions;
We use a series expansion again, and since their differential equations  are related by the transformation
\be
z_v\to z_u, \ \ \ F\to -F
\ee
we can use the same series  by simply applying the above modification.
The small $u$ behavior of $U_{nW}(u)$ can thus be written as:
\be
U_{nW}(u) = u_{n0}(W)M(W,1-z_v,-F|u) \ .
\label{eqn:UeqIn}
\ee
As we know the $U$ eigenfunctions are not normalizable, so $u_{n0}$ cannot be determined
as easily as $v_{n0}$. This is in fact the most subtle part of the whole procedure and we
will address it in its due course. However, it will be necessary to calculate $U_{nW}(u)$
for a large $u$, typically in the range of $u\sim 10^3$. This is not possible with
the series for the vicinity of the origin --- with hundred to two hundred terms in the series
one can obtain the function with a good accuracy for up to $u\sim 40$. However, the series can
be easily analytically continued to reach even very large values of $u$ as follows.

Let us assume that we have used the above series to calculate $f = M(a)$ and $p = M'(a)$ for some
$a>0$, utilizing a large number of terms (e.g. 150, to ensure sufficient accuracy). As a next step
we can obtain a series expansion of $M$ centered at $a$:
\be
M_a( W,z,F|u) = \sum_{k=0}^\infty m_k(a, f, p, W,z,F) (u - a)^k \ .
\label{eqn:MaEq}
\ee
The recursion relations for the coefficients can be obtained by inserting
into (\ref{eq:UVeqs})
  (note the we use the same master function $M$ for both $V$ and $U$ functions),
giving
\begin{align}
&m_0 = f, \nonumber \\
&m_1 = p, \nonumber \\
&m_2 =  \frac{1}{2 a} \left( \frac{F a^2 m_0}{4} -  \frac{a W m_0}{2} - z m_0 - m_1\right) , \nonumber \\
\label{eq:mcoefs}
  &m_3 =  \frac{1}{6 a}\!\left(\!\frac{F a m_0}{2}   -  \frac{W m_0}{2} +  \frac{F a^2 m_1}{4} -  \right. \\ 
&\phantom{m_3 =   \frac{1}{6 a}}   \left. \frac{a W m_1}{2} - z m_1 - 4 m_2\!\right) , \nonumber \\ 
&m_k =  \frac{1}{ k (k-1) a}\! \left(\! \frac{F m_{k-4}}{4}  + \frac{F a m_{k-3}}{2} -  \frac{W m_{k-3}}{2} +   \right.  \nonumber \\
&\phantom{s_k = } \left. \frac{F a^2 m_{k-2}}{4} -\frac{a W m_{k-2}}{2} - z m_{k-2} - (k-1)^2 m_{k-1}\right) \ . \nonumber
\end{align}
This specifies a series expansion valid around $u=a$, for the function value and its derivative at this point
set to $f$, and $p$, respectively.
  Note that this recursion scheme applies equally to both $U$ and $V$ functions since both are expressed
  in terms of the same master function $M$; the only difference is in the parameters passed to $M$.
  Obviously, this is a more complicated recursion than (\ref{eq:VcoeffEq}), because it solves a more general problem
  characterized by three additional parameters ($f,p,a$). However, this is the most important piece of the whole scheme
  because {\em it provides locally exact solutions which can be evaluated to an arbitrary precision.}

Next, one can repeat the same procedure, obtaining a new series
centered around $u=b>a$,
\begin{align}
  &M_b( W,z,F|u) = \nonumber  \\ 
  &\sum_{k=0}^\infty m_k(b, M_a(...|b) , M_a'(...|b), W,z,F) (u - b)^k \ . 
  \label{eq:Mb}
\end{align}
and keep repeating the re-expansion until reaching the $u$-region of interest. Typically, with long-double precision
and utilizing between hundred and two hundred terms, one can create a set of power-series expansions, center of each
shifted further beyond the previous. Such a representation of the function is fast to evaluate and remains
accurate for $u$ as large as several thousands.
  One could be worried that errors can accumulate by a repeated re-expansion, and they indeed could. However,
  because  one can easily increase the number of terms in the expansion, the accuracy is only limited
  by the number of bytes used in the floating-point type. So the error propagation and accumulation
  turns out to be a non-issue with the long double type (simple double type can also be used with smaller distance between patch-centers $b-a$) when
  the two values ($f$ and $p$) passed from a patch to the next are calculated to fifteen to eighteen digits.

\begin{figure}
\centering
  \includegraphics[clip,width=0.9\linewidth]{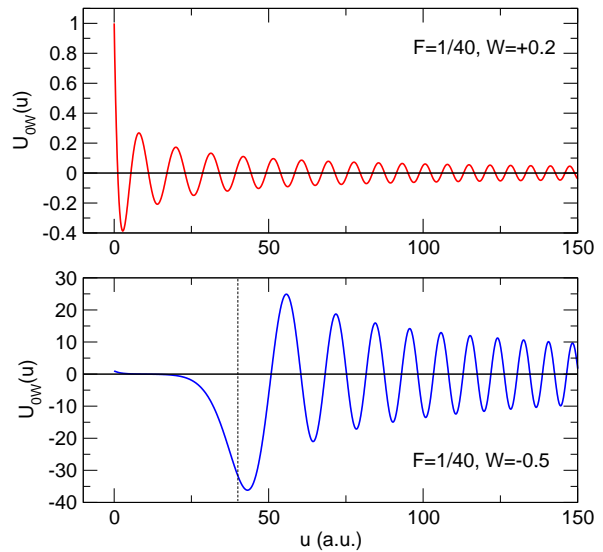}
\caption{Oscillatory behavior of $U_{nW}(u)$ for negative (bottom)
  and positive (top) energy. The dashed vertical line separates the
  classically allowed and forbidden regions.
\label{fig:Uns}
}

\end{figure}

The functional shape of $U_{nW}(u)$ is distinct from that of $V_{nW}(v)$, as should be expected for
the continuous energy spectrum. Figure~\ref{fig:Uns} shows a couple of examples, highlighting
the oscillatory nature of these functions for large argument values. While not evident on the
scale of this figure, the oscillation frequency increases toward infinity, reflecting the
fact that the particle is accelerated by the external field.

\smallskip
\noindent
\smallskip

At this point, we have a practically usable implementation of the eigenstate up to a multiplicative constant $u_{n0}$. This remaining piece is a function of $W$ and $F$,
and it carries  information crucial for the set of functions to be used as a continuum basis.
It will be obtained next.

\subsection{Large-$u$ behavior and normalization of $U_{nW}$}

In order to fix $u_{n0}$, we need to turn to the Dirac-delta normalization requirement (\ref{eq:norms}). Obviously, an asymptotic solution
for large $u$ is needed, and the strategy is to use an ansatz to separate a fast changing ``carrier'' of the wave function from its
slowly changing ``envelope''.
Looking back at the asymptotic differential equation~(\ref{eq:asymEqsU}),  one can see that the relevant solutions should behave  essentially
as Airy functions~\cite{vallee}.  The following linear combinations of Airy functions,
\be
Ci[z]^{\pm} = Bi[z]\pm i Ai[z] 
\ee
is suitable to serve as the carrier.
More concretely, our ansatz for $U$  can be written as:
\be
U\approx 
\frac{1}{2 N_U} Ci^+\left[\alpha\left(\frac{u}{2}+\frac{W}{F}\right)\right]H(u)e^{i\delta(W,F)}+
c.c.
\label{eqn:UeqnOut}
\ee
in which $N_U$ is some normalization constant,  $\alpha=-(2F)^{1/3}$, $\delta(W,F)$ is as yet undetermined phase shift, and $H(u)$ is the slowly changing
envelope of the wave function. The rationale behind this is that the envelope can be calculated
with very modest numerical effort. In fact,
its analytic asymptotics will be sufficient for many purposes.
Substituting this ansatz in the differential equation for $U$ we get:
\be
uH''
+\left(1+\alpha uR^+(u)\right)H'
+\left(z_u+\frac{\alpha}{2} uR^+(u)\right)H
=0
\label{eq:forH}
\ee
where
\be
R^+(u) =  \frac{Ci^{+'}\left[\alpha\left(\frac{u}{2}+\frac{W}{F}\right)\right]}{Ci^{+}\left[\alpha\left(\frac{u}{2}+\frac{W}{F}\right)\right]}
\ee
stands for the logarithmic derivative of $Ci^+$. Unlike the function itself, $R$ changes slowly with $u$ and as a consequence, the differential equation
for the envelope is easy to solve numerically.

Indeed, such a calculation can yield $H(u)$ which can be used in practice.
Nevertheless, here we will only utilize an asymptotic solution for $H$ valid for large $u$,
so that we will avoid numerical treatment of $H$. One reason to avoid numerical ODE solution
and favor an analytic approach is that any numerical ODE solver is designed for universal usage, and as such it can not take advantage of the concrete equation solved.
Another reason to avoid numerical ODE solution is the superior accuracy and speed of the algorithm given next.

Using the asymptotic behavior of the Airy functions, lengthy but straightforward calculations give, for $u\to\infty$,
\be
H(u) \sim \frac{1}{u^{1/2}} - \frac{2 i z_u}{F^{1/2} u} - \frac{2 z_u^2}{ F u^{3/2}} + \frac{i ( 8 z_u^3 + 4 W z_u - F )}{6 F^{3/2} u^{2}} \ldots
\label{eq:aprH}
\ee
where more terms can be calculated with some effort. 

We only need the first term in~(\ref{eq:aprH}) to obtain $N_U$, which is fixed such that the normalization condition
\be
\frac{\pi}{2}
\int
u\ U_{nW}(u)U_{nE}(u)du = \delta(W-E)
\label{eq:DDnorm}
\ee
is satisfied. This requires a calculation involving the asymptotic
behavior of $U$ that is determined by the carrier wave $Ci^+$ alone, and
leads to the normalization factor
\be
N_U = 2^{-1/3} \sqrt{\pi} F^{1/6} \ .
\label{eqn:NormU}
\ee
Having properly normalized the eigenstates to the Dirac-delta in energy, we can proceed to calculate the last remaining unknowns namely
the phase shift $\delta_n(W,F)$ and $u_{0n}(W,F)$.  The inner solution in terms of the series expansion (\ref{eqn:UeqIn}) and the outer solution expressed in (\ref{eqn:UeqnOut})
satisfy the same second-order differential equation, so in order to obtain these two unknown parameters it suffices to require that the two representations
agree at two arbitrary locations. Taking two arbitrary points $u_{1,2}$ we have a system of two equations ($i=1,2$):
\begin{multline}
u_{n0}(W,F) M_a(W,1 - z_v,-F\vert u_i)=\\
\frac{1}{2 N_U}Ci^+\left[\alpha\left(\frac{u_i}{2}+\frac{W}{F}\right)\right]H(u_i)e^{i\delta_n(W,F)}+
c.c.
\label{eq:joinio}
\end{multline}
from which $u_{n0}(W,F)$ together with $\delta_n(W,F)$ are calculated numerically. In practice, the left hand side of these
equations is obtained by the analytic continuation with the series-center $a$ in the vicinity of the  $u_1 \sim u_2$.
The advantage of the ``analytically continued'' representation of $U$ is
that the inner-outer join points $u_{1,2}$ can be taken to a very large distance form the origin
where the asymptotics of the envelope (\ref{eq:aprH}) can be used. Thus, no numerical-ODE solution of (\ref{eq:forH}) is required.



\begin{figure}[h]
  \includegraphics[clip,width=8cm]{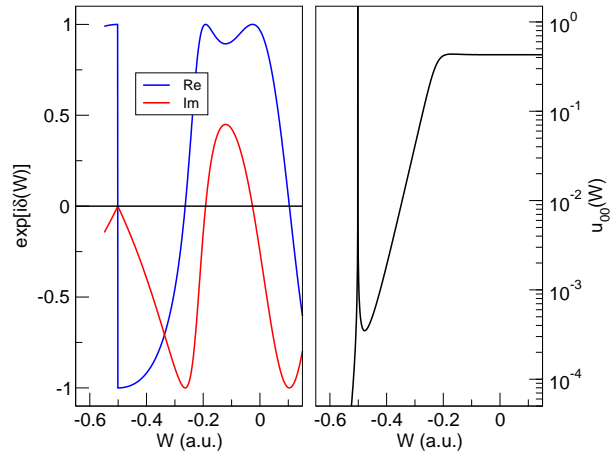}
\caption{
  Left: Phase shift $e^{i \delta_0(W)}$ as a function of energy $W$. The ``jump''
  around $W\approx -0.5$ is due to the Stark resonance born from the
  ground state. Right: $u_{00}(W)$ exhibits a sharp peak due to the same
  resonance (it is cut off in this graphics).
  \label{fig:u0delta}
}
\end{figure}

Quantities $u_{n0}(W)$ and $e^{i\delta_n(W,F)}$ are of central importance here.
They are illustrated in Fig.~\ref{fig:u0delta}.

While the phase shift is a useful quantity when one aims to calculate, for example,
the quasi-classical approximation of the wave function at large distances from the nucleus,
$u_{n0}(W)$ is necessary to finalize the calculation of the inner solution for $U_{nW}(u)$.
A most prominent feature in $u_{n0}(W)$  is the peak corresponding to the Stark resonance. This requires
that $u_{n0}(W)$ is sampled on a fine grid in the vicinity of $W\approx-1/2$. 
A fit with a simple-pole function can yield the value of the complex-valued
resonance energy accurate to a part in a million~\cite{Benassi,Jentschura}.

\section{Summary of the algorithm}

We have thus arrived at the central result of this work, which is
the algorithm for calculating the eigenstates of the Stark Coulomb problem.
The method can be summarized as follows:

\textbf{1. Fix $W$, $F$ and $n$:}\\ We start by choosing a fixed value of the external field $F$, a real energy eigenvalue
$W\in (-\infty,+\infty)$, and an integer $n$ which is the desired number of zero-crossing the wave function has along
the $v$-axis. For many experiments utilizing femtosecond optical pulses, $F$ is around $0.01 - 0.05$ in atomic units.
The relevant interval for $W$ is from about -0.6 to +0.6;

\textbf{2. Calculate $z_v(n,W,F)$:}\\ Using (\ref{eq:MVEq}) and (\ref{eq:VcoeffEq}) find numerically a value of
$z = z_v(n,W,F)$
which ensures that the function $M(W,z_v,F \vert v \to \infty) \to 0$ converges to zero at large values of $v$,
and that there are $n$ zeros of $M$.
This gives the solution to the eigenvalue problem (\ref{eq:UVeqs}).

\textbf{3. Construct $V_{nW}$:}\\
Having found $z_v(n,W,F)$, we have all necessary ingredients to calculate $V_{nW}(v)$. The remaining piece is
the normalization factor $v_{n0}(W,F)$ in (\ref{eq:VEq}) which we can calculate by integrating over
$|V_{nW}(v)|^2$ to satisfy the normalization condition (\ref{eq:Vort}).
This completes calculation of the normalized $V_{nW}(v)$ function.

\textbf{4. Construct $U_{nW}$:}\\ Setting  $z_u = 1 - z_v(n,W,F)$ according to (\ref{eq:separation}),
use (\ref{eq:MVEq}) and (\ref{eq:VcoeffEq})  and subsequently (\ref{eqn:MaEq},\ref{eq:Mb}) with (\ref{eq:mcoefs}) and
$a = k \Delta a$, $k=0,1,2\ldots$ using $\Delta a \sim 1$ on the order of unity
to calculate 
$U_{nW}(u)/u_{n0}(W) = M_a(W,z_u,-F\vert u)$. To fix the normalization, we calculate  analytically continued
series expansions (\ref{eqn:MaEq}) for $u$ up to several thousand.
Finally, $u_{n0}(W)$ is calculated by solving the system of two equations given in (\ref{eq:joinio}) using
asymptotic representation (\ref{eq:aprH}) for the envelope. As a by-product in this step, the phase-shift $\delta_n(W,F)$
is also obtained. 
This completes the calculation of $U_{nW}(u)$ satisfying the Dirac-delta normalization condition (\ref{eq:DDnorm}).

  Note that the same ``continuation method'' employing formulas (\ref{eq:Mb},\ref{eq:mcoefs}) can be also applied to functions
  $V$ as the only difference is in the input, where the sign of $F$ needs to be reversed and the
  appropriate separation constant $z_v$ or $z_u$ must be used.

\textbf{5. Obtain the energy eigenstate $\Psi_{nW}$:}\\
The eigenstate for the chosen $W$ and $n$ is given by the product in (\ref{eq:PsiUV}),
expressed as
\begin{align}
\Psi_{nW}(u,v) = &v_{n0}(W) M(W,z_v(n,W,F),F\vert v) \times \nonumber \\
&u_{n0}(W) M(W,1-z_v(n,W,F),-F\vert u)
\end{align}

\section{Implementation and usage considerations}

The method presented in this work deviates from the numerical solutions
implemented on grids (e.g.~\cite{LucKoenig1}) in that the resulting algorithm
is essentially a formula which is both simple to implement and very fast to execute.
The code can be written in less than three hundred lines of c++, and its
core for the series expansion is much smaller yet. The resulting speed
of the algorithm is such that the compute time
is not an issue. The computational complexity is comparable to that
of some special functions.

The implementation does utilize a few ``hyper-parameters'' which control the
evaluation. These include the order of the series expansion ($o_v$ and $o_u$),
the size of the patch $p_v$ and $p_u$ which is the distance between the centers
of the series-expansion components (it is $b-a$ in formula (\ref{eq:Mb})),
and the two arbitrary points $u_1$ and $u_2$ selected for joining the outer
and inner solutions.

Of course, the results must be insensitive to the choice of these values.
Thanks to the fast evaluation speed, it is straightforward to establish that
they indeed are by repeating the calculation after variation of these hyper-parameters.

To give an example of appropriate values, for our illustrations we have used
$o_v=100$ and $p_v=20$ for the $V$-solutions and $o_u=25$ with $p_u$ between one and three.
Increasing the expansion orders by fifty percent did not change the results by more than
one part in $10^{10}$ indicating that these values represent a very safe choice.

The least trivial choice of the hyper-parameters is that of $u_{1,2}$, but this is not
because they would need to be fine-tuned in any way. They merely need to be far from the
origin to make the asymptotic envelope representation (\ref{eq:aprH}) accurate. So we
used $u_{1,2}$ of the order of several thousands to ensure accuracy of the $U$-function
amplitude $u_{00}(W)$ to a part in a million or better. By moving $u_{1,2}$ further,
to $u_i\approx 10^4$, the accuracy can be further increased to a relative level of $10^{-10} - 10^{-12}$. 

The distance between the join-points
$u_2-u_1$ was varied between unity and several hundred without a significant impact on accuracy
(change in $u_{00}(W)$ less than $10^{-10}$).

\begin{figure}[h]
  \includegraphics[clip,width=0.9\linewidth]{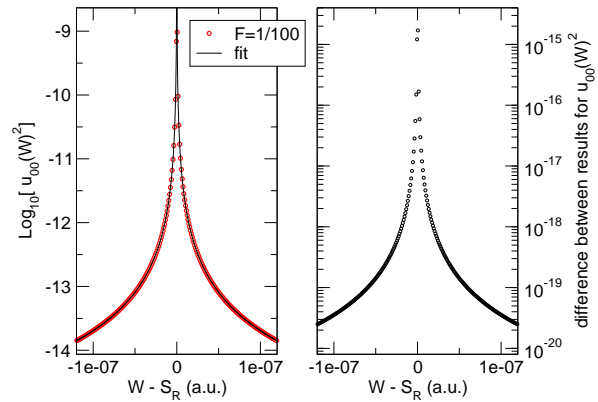}
\caption{
  Left: Behavior of $u_{00}^2(W)$ in the vicinity of the Stark resonance energy $S_R=-0.500225560455...$.
  This data allows to estimate the location of $S_R$ with an accuracy of $10^{-12}$.
  Right: Demonstration of the robustness w.r.t. choice of the join-points $u_{1,2}$.
  The plot shows the difference between results obtained with $u_1=2900,u_2=2910$ and with
  $u_1=3900,u_2=3910$. 
  \label{fig:SRcheck}
}
\end{figure}

In order to demonstrate the robustness and the accuracy of the algorithm, Fig.\ref{fig:SRcheck}
depicts data obtained for the amplitude $u_{00}(W)$ for the energy in a close neighborhood
  of the Stark resonance, which is located at $S_R=-0.500225560455...$ for $F=1/100$. We choose
  this quantity here because it is the most sensitive as it varies over many orders of magnitude
  in an extremely narrow interval of energies as shown in the left panel of the figure. The panel
  on the right depicts the difference between $u_{00}$-values obtained for two very different
  choices of $u_{1,2}$, i.e. the locations used to join the outer and inner solutions. One can see
  that despite moving $u_{1,2}$ by a thousand atomic units, the resulting numerical variation in
  $u_{00}$ is extremely small. Even as the value changes by five orders of magnitude, the relative
  error (as estimated by the difference shown) remains on the level of $10^{-6}$. It should be emphasized
  that the robustness of the  $u_{00}$-value reflects the accuracy of the wave function everywhere
  as this quantity ``connects'' the asymptotic region to the origin.

  An additional evidence for the fidelity of our algorithm is in the high accuracy of the
  Stark resonances. For the example shown in Fig.~\ref{fig:SRcheck}, we have estimated the resonance
  location $S_R$ by finding the point in the middle of the peak for higher and higher value(s) of $u_{00}$.
  We have obtained an estimate which agrees with the value from the literature~\cite{Benassi,Jentschura} to twelve digits.

\section{Resonant contribution}

We have seen that there is a potentially non-trivial $W$-dependent feature in the $u_{n0}(W)$ caused by
the Stark resonance born from the originally stable ground state. Depending on the field
it can be so narrow that it could remain unresolved, or even completely missed when $W$ is
sampled on a coarse grid. So a question arises how to deal with it, especially
for applications in time-dependent problems. Here we sketch how the
resonant contribution can be extracted and treated separately from the
continuum background.

Our previous study of toy models in one spatial dimensions suggested the functional shape
of $u_{n0}(W)$ in the vicinity of a resonance~\cite{mohammad1}.
More specifically, every resonance causes a ``step''
in the phase shift $\delta$, and in its neighborhood  we can use
\be
\exp[i \delta(W,F)] \approx  \frac{\sqrt{W- S(F)^*}}{\sqrt{W- S(F)\phantom{^*}} } \ ,
\ee
where the phase is controlled by the Stark resonance located at $S(F) = S_R(F) + i S_I(F)$.
The effect of the square root ratio is a very fast phase change with $W$ and this is because
the imaginary part $S_I$ of $S(F)$ is very small  (see Fig.~\ref{fig:u0delta}).

Incorporating this into the carrier-envelope representation of the wave function, we have
\be
u_{n0}(W) =
\frac{H(0)}{2 N_U} \mathrm{Ci}^+\left[\alpha \frac{W}{F} \right]
\sqrt{\frac{W-S(F)^*}{W-S(F)\phantom{^*}}}
 + \text{c.c.} \ .
\ee
For a sharp resonance (i.e. a weak field), the Airy combination $Ci^+$ is dominated by $Bi$, and we
can further approximate
\be
u_{n0}(W\sim S_R) =
\frac{H(0)}{2 N_U} \mathrm{Bi}\left[\alpha \frac{S_R}{F} \right] 
\sqrt{\frac{W-S_R +i S_I}{W-S_R - i S_I}}
 + \text{c.c.} \ .
\ee
where $H(0)$ is evaluated for $W=S_R$. Now we want to use an observation based on our numerical data.
Namely, we have found that $H(W=S_R,u=0)\!=\!i H_I(0)$ is purely imaginary, which leads us to
%
the following representation for the resonant part
contribution to the probability density (in energy):
\be
u_{n0}^2 \approx
\frac{\pi S_I H_I^2}{ N_U^2} \mathrm{Bi}\left[\alpha \frac{S_R}{F} \right]^2 
\frac{S_I}{\pi\left( (W-S_R)^2 + S_I^2\right)} \ .
\label{eq:resfun}
\ee
Here, the last term tends to a delta function in a weak field when $S_I\to 0$.
The fit shown as full line in Fig.~\ref{fig:SRcheck} demonstrates that  this functional shape
indeed dominates the resonant portion of $u_{00}(W)$. Symbols represent the
calculated values, and the solid lines are fits based on the above expression.
The fit was obtained with merely fifteen datums closest to the resonance. Nevertheless,
the agreement with the data remains extremely good over the whole interval shown in the
figure.
%

We have thus shown that in case the field $F$ is weak the resonance contribution to
the wave function can be accurately extracted with the help of the analytic structure
built into the carrier-envelope representation. This can present an advantage in applications,
as it offers the option to represent the spectral amplitude $A_n(W)$ as consisting of
a smooth background plus a resonant pole contribution.

\section{Illustration: Tunneling dynamics}

In general, using a continuum-energy basis to expand time-dependent wave functions is
a highly nontrivial task from the numerical point of view.
%
For instance, one of the challenges in methods utilizing discretized continuum basis sets,
is that the wave function can only be calculated within a relatively small computational box
surrounding the system. Our illustration aims to emphasize that with an accurate
algorithm to evaluate all energy eigenstates, we are free of this problem and wave functions
can be obtained even for very large distances from the origin.

We choose to consider an electron tunneling from the hydrogen atom, and calculate
its time-dependent wave function. We assume that the initial state is the zero-field
ground state wave function, and that the field is suddenly set to a constant value $F$.
Motivated by its simplicity, this is an idealized scenario previously studied in exactly
solvable one-dimensional models~\cite{mohammad1}. Here we investigate the same dynamics
in a realistic three-dimensional system.

The time-dependent solution can be expressed as a superposition of Hamiltonian eigenstates
calculated for a fixed $F$,
\be
\Psi(t) = 
\sum_n \int dW A_n(W) e^{-iWt} \Psi_{nW}  \ ,
\ee
in which the energy-representation $A_n(W)$ must be set such that the
initial condition
\be
\Psi(t=0) = \psi_G = \frac{1}{\sqrt{\pi}}e^{-(u+v)/2}
\ee
is satisfied. Using the resolution of unity in the energy space,
$\langle \Psi_{nW} \vert \Psi_{mE}\rangle = \delta_{mn}\delta(W-E)$, spectral amplitude $A_n(W)$
is obtained as an overlap integral with the initial wave function,
\begin{align}
&A_n(W) = \langle \Psi_{nW} \vert \psi_G \rangle = \\ \nonumber
&2\pi \int_0^\infty  \int_0^\infty  \frac{(u+v)}{4}
e^{-\frac{u+v}{2}} V_{nW}(v) U_{nW}(u) du dv \ .
\end{align}

\begin{figure}[!h]
  \includegraphics[clip,width=0.8\linewidth]{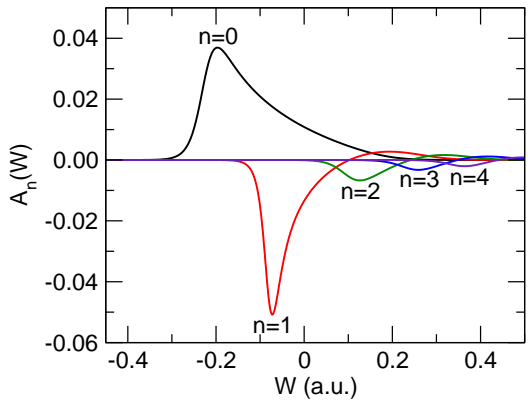}
\caption{
  Spectral amplitudes $A_n(W)$ as function of the energy $W$ for several quantum
  numbers $n$, calculated for the hydrogen-atom ground-state and external field $F=1/40$a.u.
\label{fig:An}
  }
\end{figure}

Figure~\ref{fig:An} illustrates the behavior of $A_n(W)$ for several lowest $n$s.
It reveals that the sudden application of the field excited multiple continua of
higher-energy states. 

One can see that the contribution of higher states diminishes quickly with their energy,
yet the resulting spectrum is very broad and it implies very fast evolution which we will
see shortly.

Because $A_n(W)$ is nothing but the energy-representation of the initial-time wave function,
$\vert A_n(W)\vert^2$ gives the probability distribution that the given energy is excited.
From Fig.~\ref{fig:An} one can see that the excitation of energies above $-0.4$ is small,
of the order of $10^{-4}$.
This means that almost all of the wave function remains ``concentrated'' around the
Stark-resonance peak, which is the feature
in $A_0(W)$ in the vicinity of the original ground-state energy.
This peak is so narrow that it is impossible to resolve properly in Fig.~\ref{fig:An}.

\begin{figure}[h]
  \includegraphics[clip,width=0.8\linewidth]{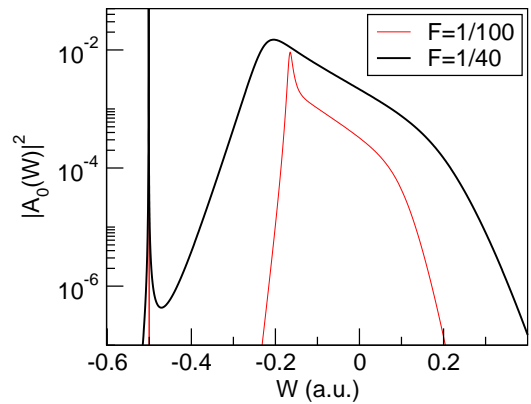}
\caption{
  Spectral amplitude $A_0(W)$ as a function of the energy $W$
  exhibits a narrow, ground-state Stark resonance. Note that on the scale of this figure
  the resonance peaks remain un-resolved and are cut-off by the plot-range.
    In contrast, the higher-energy resonances are blurred as a consequence of the
    suddenly turned-on field. The sharp feature in the curve for $F=1/100$
    is an example of the resonance ``remnant.''
\label{fig:A0log}
}
\end{figure}

Figure~\ref{fig:A0log} shows a logarithmic-scale view, and allows one to appreciate the presence of
the resonance contribution. It also indicates that two very different time scales govern the evolution
of the wave function after the field is applied.
       
The time-energy uncertainty relation suggests that the narrow resonance  peak 
gives rise to a slowly evolving part of the wave function, and the wide peak
results in a fast changing part of the wave function. The former remains
localized and very similar to a bound state, only slightly deformed by the
external field.
  The probability that the electron occupies this state can be obtained by
  fitting the resonance functional shape (\ref{eq:resfun}) to the $|A_0(W)|^2$ finely sampled
  around $W\sim S_R$. This ``amplitude'' is the survival probability of the
  system`s state after sudden exposure to the field.
  For our illustration case shown here, one obtains the survival probability of 99.9\%.
  This is the part of the wave function which remains locked in the Stark resonance.
  It gives rise to an exceedingly slow but steady leakage of the probability density
  from the vicinity of the nucleus toward infinity.
  But the survival rate of 99.9\% means that with probability of about 0.1\%,
  the electron escapes from the atom, and it occurs very fast.

\begin{figure}[h]
  \includegraphics[clip,width=0.99\linewidth]{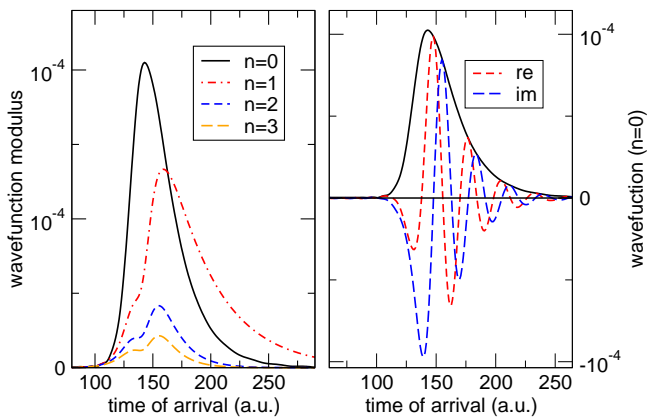}
\caption{
  Tunneled electron wave function observed at a distant location at $z=250$.
  Left: Different $n$-components arrive at the observation point in the form
  of well-defined pulses. Right: Resolved real and imaginary parts for the $n=0$
  component.
\label{fig:TOfArrivalN0}
  }
\end{figure}

This is a consequence of the non-adiabatic change in the field, and it
manifests itself as a ``pulse'' in which the electron
escapes from the atom. Figure~\ref{fig:TOfArrivalN0} illustrates this
in a ``time-of-arrival'' picture, where the evolution of the wave function
is observed at a fixed point $z=250, x=y=0$. Because the localized portion of the
wave function is exceedingly small at this distance, one can only see
a pulse for each component $n=0,1,2,3,\ldots$ as it propagates away from the
nucleus.

\begin{figure}[h]
  \includegraphics[clip,width=0.9\linewidth]{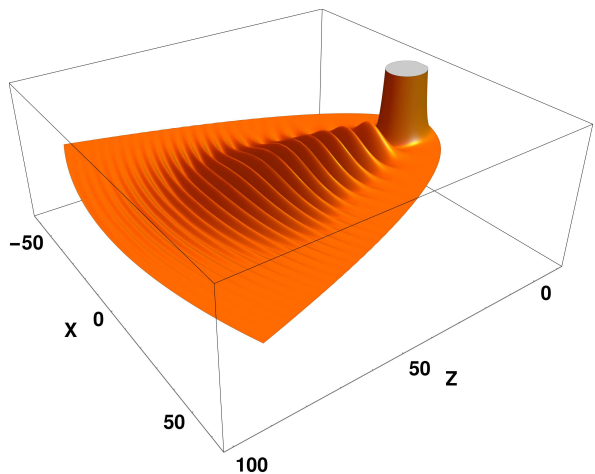}
\caption{
  Wave function snapshot taken at time $t=60$a.u. The real part of the component $n=0$ is shown here
  with the vertical plot range of $\pm0.001$ versus $x$ and $z$ in atomic units.
   The cut-off part in the center (the ``stump'')
    is the (slightly deformed by the field) wave function remaining in the metastable ground state
    which will ``survive'' for $\sim 10^{10}$ atomic units of time. In contrast, the
    waveform will propagate away from the atom very quickly, and thus contributes to the non-adiabatic
  ionization.
\label{fig:snap2dn0}
  }
\end{figure}

\begin{figure}[h]
  \includegraphics[clip,width=0.9\linewidth]{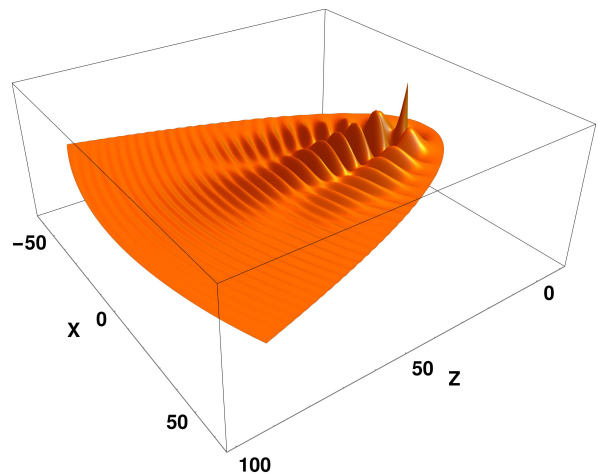}
\caption{
  Wave function snapshot taken at $t=60$a.u. The real part of the component $n=1$ is shown here
  with the vertical plot range of $\pm0.001$ versus $x$ and $z$ in atomic units.  The spatial structure indicates
  that one can`t assign a single classical trajectory to this tunneling particle.
\label{fig:snap2dn1}
  }
\end{figure}

For a snapshot-view, Fig.~\ref{fig:snap2dn0} shows the $n=0$ component depicted
at time $t=60$ after the external field was turned on. Figure~\ref{fig:snap2dn1}
shows the $n=1$ component which exhibits spatial transverse structure reflecting
directional distribution of the tunneling particles. The higher $n$ components of the
escaping wave function have correspondingly richer structure.

These figures illustrate the part of the tunneling wave function which can be
called non-adiabatic, because it is caused by the sudden turn on of the field.
Given the distance from the nucleus, and the accuracy required, this
would be extremely difficult to evaluate with standard numerical methods.

{\revcolor{red} 
  \circled{II}
  It is worth noting that our wave function, as shown in Fig.~\ref{fig:snap2dn0} contains both the
  eventually freed part and the part which is bound to the nucleus for a very long time and we do not
  explicitly distinguish between the two in the formulas presented. However, at larger distances 
  the part that is bound to the nucleus is exponentially small in comparison to the ``freed" part,
  and what one obtains far from the nucleus is for all practical purposes the part of the wave function
  which represents the ``freed'' electron.
}


  Our example is therefore relevant for the ongoing debate concerning the question of the tunneling
  time. More specifically, there is no universal agreement about the time an electron spends
  while tunneling from an atom under the influence of an external electric field
  (a brief discussion of different schools of thought can be found in \cite{mohammad1}). 
  {\revcolor{red}
  \circled{II}
   In this context, our example
  is the first exact solution in a three-dimensional system that suggests the tunneling time and velocity might not be zero.
  In this model there is no single well-defined tunnel exit, and therefore  the notion of
  ``tunneling time'' is also ambiguous.
}
  As it can be seen from Fig.~9 and 10, we can also conclude that in this   three-dimensional case we cannot
  ascribe a single classical trajectory to the tunneling portion of the wave function.
%
  The tunneling dynamics continues to be quantum in nature even beyond the emergence
  of the particle from the classically forbidden region. Consequently, proper quantum
  description should be preferred also in the outer regions, and our
  representation of the Stark-Hydrogen solutions represent an ideal tool for such
  an application.
  {\revcolor{red}
     \circled{II}
    For example, the same kind of calculations as above only at larger distances can be used to create data for
    the back-propagation method to find a family of classical trajectories
    (see e.g. Ref.\cite{PhysRevA.98.013411} for various classifications of the exit and initial velocity).

    In relation to the debate about the nature of the tunneling time,
    we have to admit that the present model does not yet completely reflect a typical experiment since such
    experiments are done using a circularly polarized (or ``almost" circularly polarized)
    laser beam.~\cite{SainadhTunnelTime,Kellerstreak} Moreover, our sudden excitation drives the
  system into a more non-adiabatic regime than an optical pulse with a smoothly varying intensity envelope would.
  Thus, it should be interesting to generalize our model to one which reflects the  circular polarization
  of the driving field. The Stark-Hydrogen states could serve as a suitable point of departure
  to describe the ionization and tunneling dynamics in a reference frame co-rotating with the electric field.
}

  Another currently relevant physics question concerns strong-field ionization in a
  long-wavelength (i.e. slowly evolving) optical field~\cite{Yudin}. We have seen
  that  there are two qualitatively different channels for ionization.
  Besides the well-known adiabatic tunneling ionization which is mediated by the
  Stark resonance~\cite{kolesik:14,SSSIvsMESA,StarkDFT,Bahl:17},
  there is also ionization current in the form of a short-duration ``pulse.''
  This is the reaction of the system to the increment of the  external field, so it is
  non-adiabatic in nature. Post-adiabatic corrections are  universally believed to become
  negligible for a very slowly evolving external fields, but accurate evaluations are
  not available at this time. Having seen here how strong the non-adiabatic contribution
  is in our case, a detailed investigation should be of great interest.
  The technique used in our example can be generalized
  to evalaute the relative strength of the adiabatic and post-adiabatic ionization yields
  in slowly evolving external fields.

\section{Conclusion}

We have presented an accurate, approximation-free algorithm to calculate
the continuum-energy eigenstates of the Stark-Hydrogen problem for a particle
subject to Coulomb and a homogeneous electric fields.

Importantly for future applications, the algorithm presented in this work
takes advantage of the analytic properties of the problem. In particular,
the wave functions are represented with the help of a ``carrier wave'' which
captures the tunneling portion of the wave function. This is then modified by
an ``envelope,'' or slowly changing function, which is relatively easy to find
numerically, or analytically in the form of an asymptotic expansion.
This representation of the wave functions can be useful in
constructing quasi-classical solutions for electrons escaping from
single-charged quantum systems.

Complementary to the carrier-envelope representation, we have also
designed an algorithm suitable for small and intermediate (up to a few thousands
of atomic units) distances from the
nucleus. This approach is based on a series expansion ``analytically continued''
away from the origin, and it can be used to study the properties and
eventually the temporal dynamics of the electrons driven by strong
long-wavelength optical fields.
 
We have illustrated the usage of the continuum-energy basis to expand
a time-dependent wave function of a tunneling particle after  an excitation
due to a sudden turn-on of an external field. Application to a general case with
time-dependent $F(t)$ requires evaluation of the spectral amplitude
$A_{nW}(t)$ which, too, depends on time. This generalization will be addressed
elsewhere. 

To the best of our knowledge, ours is the first practically usable
state-expansion method utilizing a continuous energy basis for a realistic
three-dimensional quantum system. Given that the superposition of the Coulomb
and homogeneous fields appears in many situations, we trust that the capability to
treat these nontrivial quantum states without approximations will prove useful
in various situations, including for example strong-field ionization of atoms and molecules,
and high-harmonic generation.

\bigskip
\noindent
{\bf Acknowledgments}\\
This material is based upon work supported by the Air Force Office of Scientific
Research under award number FA9550-18-1-0183.

\section{Appendix: Normalization}


\bigskip
\noindent

The normalization condition we seek to satisfy reads
\begin{align}
\nonumber
2\pi\int \frac{u+v}{4} 
& V_{nW}(v) V_{mE}(v)
U_{nW}(u) U_{mE}(u)
dvdu\\ 
&=
\delta_{nm}\delta(W-E) \ ,
\label{app:complete}
\end{align}
and expands into two contributions which we will refer to as $u$- and $v$-terms:
\begin{align}
  I_u\! &=\!\frac{\pi}{2} \int  V_{nW}(v) V_{mE}(v) dv \int u U_{nW}(u) U_{mE}(u) du \\
  I_v\! &=\!\frac{\pi}{2} \int  v V_{nW}(v) V_{mE}(v) dv \int  U_{nW}(u) U_{mE}(u) du ,
\label{app:uterms}
\end{align}
together giving rise to the delta function in energy like so
\begin{equation}
I_u + I_v = \delta_{mn} \delta(W-E) \ .
\end{equation}
Let us look first at the integrals over $v$. Because the $V$-functions are integrable,
they can be normalized to unity. It can also be shown in a standard way that they are
orthogonal to each other, so we can assume
\begin{equation}
  \int_0^\infty  V_{nE}(v) V_{mE}(v) dv = \delta_{mn} \ .
  \label{eqn:normV}
\end{equation}
For the other integral, it turns out that the non-diagonal part will not be needed,
and one can show that the diagonal portion can be simplified as
\begin{equation}
  \int  V_{nW}(v) v V_{nW}(v) dv = -2 z_v'\ ,
  \label{eqn:VvV}
\end{equation}
where prime denotes partial derivative with respect to $W$.
To obtain this result, one takes the ``expectation value'' equation for  $h_v$,
   \begin{equation}
\int_0^\infty V_{nW}(v) \hat h_v V_{nW}(v) dv  = z_v(n,W) \ ,
   \end{equation}
   differentiates on both sides w.r.t. $W$, and uses the fact that the normalization
   is fixed as in (\ref{eqn:normV}).

Now we turn to the integrals over $u$. With the normalization of $V$ accounted for,
the $I_u$ term  turns into
\be
I_u = \delta_{mn} \frac{\pi}{2} \int_0^\infty  U_{nW}(u)\  u\ U_{nE}(u) du  \ .
\ee
The relevant contribution to this integral is from the region of large $u$,
where we use the asymptotic behavior of the $U$-functions,
\be
U = \frac{1}{2 N_U} \mathrm{Ci}^+\left[\alpha \left( \frac{u}{2} + \frac{W}{F}\right) \right] H(u)
e^{i \delta(W,F)} + \text{c.c.} \ ,
\label{eqn:outerU}
\ee
where $H$ is the envelope which behaves as $H\sim u^{-1/2}$ at infinity
and thus cancels $u$ in the integrand. The integral over large values of $u$ tends to
(we leave out index $n$ in the integrand)
\begin{widetext}
  \be
  I_u = 
\delta_{mn} \frac{\pi}{2} \int_0^\infty
\left( \frac{1}{2 N_U} \mathrm{Ci}^+\left[\alpha \left( \frac{u}{2} + \frac{W}{F}\right) \right] e^{i \delta(W,F)} + \text{c.c.} \right)
\left( \frac{1}{2 N_U} \mathrm{Ci}^+\left[\alpha \left( \frac{u}{2} + \frac{E}{F}\right) \right] e^{i \delta(E,F)} + \text{c.c.} \right)
du  \ .
\ee
Eliminating terms which oscillate even for $E=W$ and do not contribute to the delta-function one gets
\be
I_u = 
\delta_{mn} \frac{\pi}{8 N_U^2} \int_0^\infty
\left(
\mathrm{Ci}^+\left[\alpha \left( \frac{u}{2} + \frac{W}{F}\right) \right] e^{+i \delta(W,F)}
\mathrm{Ci}^-\left[\alpha \left( \frac{u}{2} + \frac{E}{F}\right) \right] e^{-i \delta(E,F)}
+ \text{c.c.} \right)
du  \ .
\ee
Using the leading term in the asymptotic expansion of the Airy functions, one obtains
\be
I_u =
\delta_{mn} \frac{1}{4\ 2^{2/3} F^{1/6} N_U^2}\int_0^\infty \left( \exp[-i \frac{u^{1/2} (E-W)}{F^{1/2}} ] e^{i \delta(W,F) - i \delta(E,F)}  + c.c.\right)
\frac{du}{u^{1/2}}  \ .
\ee
\end{widetext}
Recall that the phase-shift terms $\delta(.,F)$ belong to the same $m=n$, and because they cancel for $W\to E$ they can be dropped.
After substitution $k=u^{1/2} F^{-1/2}$ and representing the complex conjugate term as an integral over $-k$, we obtain
\be
I_u=
\delta_{mn}\frac{2 F^{1/2}}{4\ 2^{2/3} F^{1/6} N_U^2}\!\!\int_{-\infty}^{+\infty}\!\!\!\!\!  \exp[-i k (E\!-\!W) ] dk 
\ee
or
\be
I_u = \frac{ F^{1/3} \pi}{ 2^{2/3}  N_U^2}\delta(E-W)\delta_{mn}  \ .
\label{eqn:Iu}
\ee
Next we  turn to the $v$-term given by the integral
  \begin{align}
    I_v &= \frac{\pi}{2} \int_0^\infty\!\! v V_{nW} V_{mE}  dv \int_0^\infty\!\! U_{nW}(u) U_{mE}(u) du\nonumber \\ &\equiv V^{(1)}_{mn}(W,E)\ I_{mn}(W,E) \ .
  \end{align}
  We will see that only the diagonal portion of $V^{(1)}_{mn}(W,E)$ plays a role, because the second term
  turns out proportional to $\delta_{mn}$. It requires to evaluate
  \begin{equation}
    I_{mn}(W,E) = \frac{\pi}{2} \!\! \int_0^\infty\!\! U_{nW}(u) U_{mE}(u) du \ ,
  \end{equation}
  which is a distribution because we deal with a continuous spectrum. Next we show that for $m\ne n$
  the above equals zero in the distribution sense. We start by rewriting it equivalently like so
  \begin{equation}
    I_{mn}\!\!  =\!\!  \frac{\pi}{2} \frac{[z_u(m,E)- z_u(n,W)]}{[z_u(m,E)- z_u(n,W)]}\!\! \int_0^\infty\!\! U_{nW}(u) U_{mE}(u) du 
  \end{equation}
  in  which we use the fact that $z_u$ are the eigenvalues of the differential operator $\hat h_u$ to obtain
  \begin{align}
    & [z_u(m,E)- z_u(n,W)] I_{mn}  =  \\
    &\frac{\pi}{2} \int_0^\infty\!\! \left[ U_{nW}(u) \hat h_{uE} U_{mE}(u) -  U_{mE}(u) \hat h_{uW} U_{nW}(u)\right] du \ ,
    \nonumber
  \end{align}
  where we have used additional indices $W,E$ in $\hat h_u$ to indicate their corresponding energy parameters.
  The parts of $h_u$ that are proportional to $F$ cancel out and one is left with
  \begin{widetext}
  \begin{align}
    & [z_u(m,E)- z_u(n,W)] I_{mn}  =  \nonumber \\
    &\frac{\pi}{2} \int_0^\infty\!\! \left[ U_{nW}(u) (-1)\partial_u ( u \partial_u   U_{mE}(u)) -  U_{mE}(u) (-1) \partial_u (u \partial_u  U_{nW}(u))\right] du +\nonumber \\
        &\frac{\pi}{2} \int_0^\infty\!\! \left[ U_{nW}(u) \frac{(-E) u}{2}   U_{mE}(u) -  U_{mE}(u) \frac{(-W) u}{2}  U_{nW}(u)\right] du \ ,
  \end{align}
\end{widetext}
One can integrate by parts twice  in the first term on the right-hand side to get
\begin{equation}
I_{mn} \equiv  I_{mn}^{(S)} + I_{mn}^{(V)} \ ,    
\end{equation}
where the ``surface term'' reads
  \begin{align}
 I_{mn}^{(S)} &= \\
 -&\frac{\pi \left\vert u  U_{nW}(u) \partial_u U_{mE}(u) -  u U_{mE}(u) \partial_u U_{nW}(u) \right\vert_0^\infty }{2 [z_u(m,E)- z_u(n,W)]}
\nonumber
  \end{align}
and the ``volume contribution'' is
  \begin{align}
  I_{mn}^{(V)} &= \\
  -&\frac{(E-W)}{2 [z_u(m,E)- z_u(n,W)]  } \frac{\pi}{2} \int_0^\infty\!\!  U_{nW}(u) u U_{mE}(u) du
  \nonumber .
  \end{align}

  This is the point where one needs to  use the specific properties of the $U$-functions. When $W\ne E$ both terms on the
  right-hand size oscillate themselves to a distribution-zero.  As a consequence,   $I_{m\ne n}(W, E)$ is zero in the distribution
  sense, meaning that $I_{m\ne n}$ acting on  an arbitrary smooth test function $T(E)$  yields zero:
  \begin{equation}
\int I_{m\ne n}(W, E) T(E) d E = 0 \ .
    \end{equation}
  However, this argument does not hold for $m=n$ and $W\to E$. We have to investigate such a case more closely,
  starting with the ``volume'' term. We know from (\ref{eqn:Iu}) that the integral part is proportional to the Dirac-delta
  in energy, so the prefactor can be replaced by its limit for $W\to E$,
  \begin{equation}
  I_{nn}^{(V)} = 
  -\frac{1}{2 z_u'(n,W)  } \frac{\pi}{2} \int_0^\infty\!\!  U_{nW}(u) u U_{nE}(u) du \ .
  \end{equation}
  As a consequence of (\ref{eqn:VvV}) and because $z_v' = -z_u'$, the contribution to $I_v$ simplifies to
  \begin{equation}
  V^{(1)}_{nn}(W,W) I_{nn}^{(V)} = -\frac{\pi}{2} \int_0^\infty\!\!  U_{nW}(u) u U_{nE}(u) du \ ,
  \end{equation}
which, interestingly, exactly cancels $I_u$ contribution to the normalization integral.

The surviving contribution is therefore given by the ``surface terms'' $ V^{(1)}_{mn} I_{mn}^{(S)}$.
  Because functions $U$ are regular in the vicinity of the origin,
  the lower bound does not contribute. For the upper boundary term ($u\to\infty$)
  the envelope $H\sim u^{-1/2}$ cancels the factor of $u$, and we end up with a bounded
  function which oscillates faster and faster for large $u$. More precisely, utilizing the
  asymptotic expansion of the Airy functions, we obtain
  \begin{align}
    &I_{mn}^{(S)}=\\  &\frac{-\alpha}{4 N_U^2} \lim_{u\to\infty}\frac{ \sin[ \delta_n(W)\!-\!\delta_m(E)\! +\! (E-W)  (u/F)^{1/2}  ] }{ z_u(m,E)\!-\!z_u(n,W) }
    + \ldots
    \nonumber
  \end{align}
  When $m\ne n$, the denominator remains finite even for $W\to E$, and as $u\to\infty$ the function oscillates
  ``infinitely fast.'' As such, it vanishes in the distribution sense.
  Let us consider the diagonal case $n=m$ 
  \begin{align}
     &I_{nn}^{(S)} =\\  &\frac{-\alpha}{4 N_U^2}\! \lim_{u\to\infty}\frac{ \sin[ \delta_n(W)\!\!-\!\!\delta_n(E)\! +\! (E-W) (u/F)^{1/2}  ] }{ z_u(n,E)- z_u(n,W) } \ .
\nonumber
  \end{align}
  The relevant  behavior is of $W\sim E$, so expanding accordingly we get
   \begin{align}
     &I_{nn}^{(S)} = \nonumber \\
     &\frac{-\alpha}{4 N_U^2} \lim_{u\to\infty}\frac{  \sin[ (E-W) (u/F)^{1/2}  ] }{ z_u'(n,W) (E - W) } \nonumber \\
     =  &\frac{-\alpha\pi}{4 N_U^2 z_u'(n,W)} \delta(W-E) \ .
  \end{align}
Once again, the derivative of the separation constant $z_u'$ gets canceled by the integral over $v$ and the
   contribution to the whole $I_v$ is:
   \begin{equation}
 V^{(1)}_{nn} I_{nn}^{(S)}   =  \frac{ F^{1/3} \pi}{2^{2/3} N_U^2}\delta(W-E) \delta_{mn} \ .
   \end{equation}
   In summary, the normalization integral gives three contributions which are equal in their absolute values
   while one of them is negative. In effect the $u$-term remains, while the $v$-term vanishes (as a distribution).
   As a result the normalization to the Dirac-delta in energy requires
   to set $N_U$ such that
   \begin{equation}
\frac{ F^{1/3} \pi}{2^{2/3} N_U^2}\delta(W-E) \delta_{mn} = \delta_{mn} \delta(W-E)\ ,
   \end{equation}
   or
   \begin{equation}
     N_U = 2^{-1/3} F^{1/6} \sqrt{\pi} .
   \end{equation}
   With the normalization factor set like this, it can be shown that the total probability outflow
   in the outgoing part of the wave function is equal to $1/2\pi$ independently of the energy,
   as it is expected to be~\cite{Landau} (\textsection 21.)


%

\end{document}